\documentclass[10pt, oneside]{article}
\usepackage{Jcappub}
\usepackage{geometry}                		
\pdfoutput=1
\usepackage{graphicx}			
\usepackage{bm}
\usepackage{array}

\newcommand{\eqb}{\begin{equation}}			
\newcommand{\eqe}{\end{equation}}			
\newcommand{\fs}[2]{#1/#2} 					
\newcommand{\fb}[2]{\frac{#1}{#2}} 			
\renewcommand{\d}{\mathrm{d}} 				
\newcommand{\de}{\partial} 					
\newcommand{\di}{\,\mathrm{d}} 				
\newcommand{\D}{\,\mathrm{\Delta}} 			
\renewcommand{\(}{\left(}				
\renewcommand{\)}{\right)}				

\newcommand{\fonttherm}{\mathnormal}		
\newcommand{\xT}{\fonttherm{T}}				
\newcommand{\xTL}{\fonttherm{T_{L}}}		
\newcommand{\xTV}{\fonttherm{T_{V}}}		
\newcommand{\xlam}{\fonttherm{\lambda}}		
\newcommand{\xrhoL}{\fonttherm{\rho_{L}}}	
\newcommand{\xrhoV}{\fonttherm{\rho_{V}}}	
\newcommand{\xp}{\fonttherm{p}}				
\newcommand{\xpL}{\fonttherm{p_{L}}}		
\newcommand{\xpV}{\fonttherm{p_{V}}}		
\newcommand{\xpb}{\fonttherm{p_{b}}}		
\newcommand{\xsig}{\fonttherm{\sigma}}		
\newcommand{\xdelR}{\fonttherm{\delta_{R}}}		
\newcommand{\xRc}{\fonttherm{R_{c}}}		
\newcommand{\xRcstar}{\fonttherm{R_{c}^{*}}}		
\newcommand{\xdelE}{\fonttherm{\delta_{E}}}		
\newcommand{\xEc}{\fonttherm{E_{c}}}		
\newcommand{\xEcstar}{\fonttherm{E_{c}^{*}}}
\newcommand{\xV}{\fonttherm{V}}				
\newcommand{\xVc}{\fonttherm{V_{c}}}		
\newcommand{\xVliq}{\fonttherm{V_\text{liq}}}	
\newcommand{\xU}{\fonttherm{U}}					
\newcommand{\xUvol}{\fonttherm{U_\text{vol}}}	
\newcommand{\xUsurf}{\fonttherm{U_\text{surf}}}	
\newcommand{\xWexp}{\fonttherm{W_\text{exp}}}	
\newcommand{\xWvol}{\fonttherm{W_\text{vol}}}	
\newcommand{\xWsurf}{\fonttherm{W_\text{surf}}}	
\newcommand{\xWb}{\fonttherm{W_{b}}}			
\newcommand{\xvr}{\fonttherm{v_{r}}}			
\newcommand{\xQevap}{\fonttherm{Q_\text{evap}}}	
\newcommand{\xH}{\fonttherm{H}}					
\newcommand{\xHvol}{\fonttherm{H_\text{vol}}}	
\newcommand{\xHsurf}{\fonttherm{H_\text{surf}}}	
\newcommand{\xG}{\fonttherm{G}}					
\newcommand{\xGprime}{\fonttherm{G'}}			
\newcommand{\xGvol}{\fonttherm{G_\text{vol}}}	
\newcommand{\xGsurf}{\fonttherm{G_\text{surf}}}	
\newcommand{\xGL}{\fonttherm{G_L}}	
\newcommand{\xGV}{\fonttherm{G_V}}	
\newcommand{\xr}{\fonttherm{r}}					
\newcommand{\xF}{\fonttherm{F}}					
\newcommand{\xFsurf}{\fonttherm{F_\text{surf}}}	
\newcommand{\xA}{\fonttherm{A}}					
\newcommand{\xS}{\fonttherm{S}}					
\newcommand{\xSvol}{\fonttherm{S_\text{vol}}}	
\newcommand{\xSsurf}{\fonttherm{S_\text{surf}}}	

\title{\boldmath On the critical energy required for homogeneous nucleation in bubble chambers employed in dark matter searches}

\author[a]{G.~Bruno,}
\author[b,c]{N.~Burgio,}
\author[d,c,1]{M.~Corcione,\note{Corresponding author.}} 
\author[d]{L.~Cretara,} 
\author[d]{M.~Frullini,} 
\author[a,e]{W.~Fulgione,}
\author[d,f]{L.~Manara,} 
\author[d,c]{A.~Quintino,} 
\author[b]{A.~Santagata} 
\author[f,g]{and L.~Zanotti} %
\emailAdd{massimo.corcione@uniroma1.it}
\emailAdd{walter.fulgione@lngs.infn.it}

\affiliation[a]{INFN, LNGS, Via G. Acitelli 22, 67100 Assergi (L'Aquila), Italy}
\affiliation[b]{ENEA, Centro Ricerche Casaccia, Via Anguillarese 301, S. Maria di Galeria, 00123 Rome, Italy}
\affiliation[c]{INFN, Sezione di Roma La Sapienza, P.le Aldo Moro 2, 00185 Rome, Italy}
\affiliation[d]{DIAEE, Sapienza Universit\'a di Roma, Via Eudossiana 18, 00184 Rome, Italy}
\affiliation[e]{INAF, Osservatorio Astrofisico di Torino, Via Osservatorio, 30, 10025 Turin, Pino Torinese, Italy}
\affiliation[f]{INFN, Sezione di Milano-Bicocca, P.za della Scienza 3, 20126 Milan, Italy}
\affiliation[g]{Dip. di Fisica, Universit\`a di Milano Bicocca, P.za della Scienza 3, 20126 Milan, Italy}

\begin{document}
\abstract{
Two equations for the calculation of the critical energy required for homogeneous nucleation in a superheated liquid, and the related critical radius of the nucleated vapour bubble, are obtained, the former by the direct application of the first law of thermodynamics, the latter by considering that the bubble formation implies the overcoming of a barrier of the free enthalpy potential. Comparisons with the currently used relationships demonstrate that the sensitivity of the bubble chambers employed in dark matter searches can be sometimes notably overestimated.
}

\maketitle
\flushbottom

\section{Introduction}
Bubble chambers using superheated liquids have been widely employed in high-energy physics for several decades after the invention of Glaser dated back to 1952 \cite{bib:1}. Recently, variants of such detectors are exploited in the search for dark matter in the form of Weakly Interacting Massive Particles (WIMPs), the main difference from the standard bubble chambers being the fact that the target liquid is continuously maintained in the metastable superheated state, instead of for just a few milliseconds \cite{bib:2,bib:3,bib:4,bib:5,bib:6,bib:7}.

In both applications, bubble nucleation is the result of a highly localized deposition of at least the minimum amount of energy required for the formation of a bubble of critical size, as postulated by Seitz in his "thermal spike" theory \cite{bib:8}, which is the model currently accepted as the best explanation available for radiation-induced nucleation in superheated liquids. The minimum amount of energy to be released as a thermal spike to produce a bubble nucleation, typically called critical energy, is generally expressed as the sum of a number of terms, this number varying with the assumptions made by each investigator. Moreover, also the value of the critical bubble radius, which enters directly into the calculation of the critical energy, depends on the assumptions made for its evaluation. Indeed, very often the theoretical values of the critical energy, i.e., the thermodynamic energy thresholds, are lower, sometimes drastically, than the corresponding experimental values, which can result in an overestimation of the bubble chamber sensitivity. On the other hand, the relatively low threshold needed for WIMP-recoil detection asks to be the most accurate as possible in the prediction of the critical energy required for bubble nucleation.

In this general framework, a reasoned review of the critical energy equations readily available in the literature, and the related expressions of the critical bubble radius, is carried out. A pair of relationships for the determination of the critical energy and bubble radius are then proposed and discussed.

\section{Critical energy for bubble nucleation}
\begin{figure}[t!]
\centering
\includegraphics[width=3.5717in,height=2.6181in]{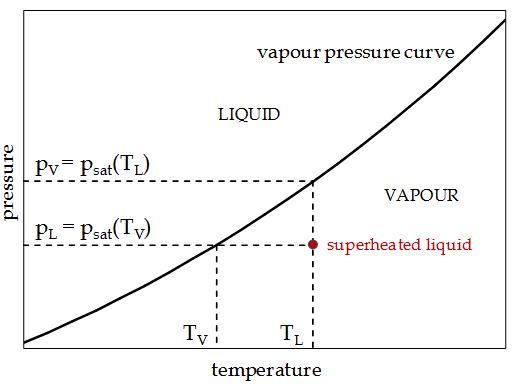} 
\caption{$\xp\xT$ phase diagram for the liquid and vapour regions}
\label{fig:1}
\end{figure}

A liquid at temperature $\xTL$ and pressure $\xpL$ is called superheated when $\xTL$ is higher than the saturation temperature $\xTV$ at pressure $\xpL$, or, that is the same, $\xpL$ is lower than the saturation pressure $\xpV$ at temperature $\xTL$, as shown in the $\xp\xT$ phase diagram depicted in figure \ref{fig:1}, in which the saturation line separating the vapour and liquid single-phase regions represents the two-phase liquid-vapour region. It can be seen that for each saturation pressure there is a unique saturation temperature and vice versa, their correspondence being described by the Clapeyron-Clausius equation
\eqb
\fb{\d\xp}{\d\xT}=\fb{\xrhoV\xlam}{\xT(1-\fs{\xrhoV}{\xrhoL})}, \label{eq:1}
\eqe
where $\xlam$ is the latent heat of vaporization, and $\xrhoL$ and $\xrhoV$ are the mass densities of the saturated liquid and vapour phases.

Notice that, strictly speaking, the metastable liquid state of coordinates $(\xTL,\xpL)$, which apparently falls in the vapour region, could not be displayed in the $\xp\xT$ phase diagram, wherein only stable equilibrium states can be represented. Of course, the degree of metastability of the superheated liquid can be expressed either in terms of superheat, $\D\xT=\xTL-\xTV$, or in terms of underpressure, $\D\xp=\xpV-\xpL$.

In a bubble chamber in which the sensitive liquid is kept superheated at temperature $\xTL$ and pressure $\xpL$, if enough energy is deposited into the liquid, the formation of a critically-sized vapour bubble occurs, its radius $\xRc$ being given by the Young-Laplace relation
\eqb
\xRc=\fb{2\xsig}{\xpb-\xpL}, \label{eq:2}
\eqe
where $\xsig$ is the surface tension of the liquid, and $\xpb$ is the pressure inside the bubble.
The critical energy $\xEc$ required for bubble nucleation has been the subject of a number of studies conducted in the past, each leading to an expression composed of different terms. The terms commonly included in the critical energy equation are the energy required to vaporize the mass of liquid involved in the phase change and the energy required to form the bubble surface. In most formulations an expansion term is added to account for the expansion work tranferred from the vapour bubble to the surrounding liquid during the vapour bubble formation, while less frequently a term accounting for the kinetic energy imparted by the expanding vapour bubble to the surrounding liquid is also incorporated. A list of the originally proposed equations are summarized in table \ref{tab:1}, in which $\xvr$ denotes the radial velocity of expansion of the bubble surface, whereas all the other variables have already been defined earlier throughout the text.

\begin{table}[!t] 
\caption{Terms in the $\xEc$ equation proposed by different authors}
\label{tab:1}
\vspace{0.5cm}
\centering
{
\renewcommand{\arraystretch}{1}
\begin{tabular}{|m{3.6cm}|m{1.9cm}|m{2.8cm}|m{2.9cm}|m{2.0cm}|}
\hline
\center{{\footnotesize{Author

 }}}& 
\center{{\footnotesize{Vaporization

} }}& 
\center{{\footnotesize{Surface formation 

}}}& 
\center{{\footnotesize{Expansion} 

}}&
{{\footnotesize{$~~~~$Kinetic}}} \\

\hline
\hline

{\footnotesize{
\center{Pless and Plano \cite{bib:9} }
}} & 
$\fb{4}{3}\pi\xRc^{3}\xrhoV\xlam$ & 
$4\pi\xRc^{2}\xsig$ & 
$\fb{4}{3}\pi\xRc^{3}\xpL$ & 
$-$
 \\
\hline
{\footnotesize{
Seitz \cite{bib:8}; 

see also \cite{bib:15,bib:16}
}}& 
$\fb{4}{3}\pi\xRc^{3}\xrhoV\xlam$ & 
$4\pi\xRc^{2}\xsig$ & 
$-$ & 
$-$ \\ 
\hline
{\footnotesize{
Bugg \cite{bib:10}; 

see also \cite{bib:3,bib:17,bib:18,bib:19,bib:20,bib:21,bib:22}
}}& 
$\fb{4}{3}\pi\xRc^{3}\xrhoV\xlam$ & 
$4\pi\xRc^{2}(\xsig-\fb{\d\xsig}{\d\xT}\xTL)$ & 
$-\fb{4}{3}\pi\xRc^{3}(\xpV-\xpL)$ & 
$-$ \\
\hline
{\footnotesize{
\center{Norman and Spiegler 
\cite{bib:11}
}}} & 
$\fb{4}{3}\pi\xRc^{3}\xrhoV\xlam$ & 
$4\pi\xRc^{2}(\xsig-\fb{\d\xsig}{\d\xT}\xTL)$ & 
$-$ & 
$2\pi\xrhoL\xRc^{3}\xvr^{2}$ \\
\hline
{\footnotesize{
\center{Tenner \cite{bib:12}} 
}}& 
$\fb{4}{3}\pi\xRc^{3}\xrhoV\xlam$ & 
$4\pi\xRc^{2}(\xsig-\fb{\d\xsig}{\d\xT}\xTL)$ & 
$\fb{4}{3}\pi\xRc^{3}(1-\fb{\xrhoV}{\xrhoL})\xpL$ & 
$-$ \\
\hline
{\footnotesize{
Peyrou \cite{bib:13}; 

see also \cite{bib:23,bib:24,bib:25,bib:26,bib:27}
}}& 
$\fb{4}{3}\pi\xRc^{3}\xrhoV\xlam$ & 
$4\pi\xRc^{2}(\xsig-\fb{\d\xsig}{\d\xT}\xTL)$ & 
$\fb{4}{3}\pi\xRc^{3}\xpL$ & 
$-$ \\ 
\hline
{\footnotesize{
Bell et al. \cite{bib:14}; 

see also \cite{bib:28,bib:29,bib:30}
}}& 
$\fb{4}{3}\pi\xRc^{3}\xrhoV\xlam$ & 
$4\pi\xRc^{2}\xsig$ & 
$-\fb{4}{3}\pi\xRc^{3}(\xpV-\xpL)$ & 
$2\pi\xrhoL\xRc^{3}\xvr^{2}$ \\
\hline
\hline
\end{tabular}
}
\end{table}

Actually, the critical energy is completely described by two terms: the vaporization term, and the surface formation term expressed in the form first introduced by Bugg \cite{bib:10}. In fact, based on the first law of thermodynamics, the heat injection required to nucleate a critical bubble, i.e., the critical energy $\xEc$, is given by the sum of the internal energy variation, $\D\xU$, and the expansion work $\xWexp$ transferred from the vapour bubble to the surrounding liquid during the vapour bubble formation
\eqb
\xEc=\D\xU+\xWexp. \label{eq:3}
\eqe

On the other hand, once the vapour bubble is thought as composed of its bulk volume and the interfacial region, conventionally assumed to have no thickness and thus represented by the mathematical surface of the bubble, $\D\xU$ can be written as the sum of a volume term, $\D\xUvol$, and a surface term, $\D\xUsurf$, giving
\eqb
\xEc=\D\xUvol+\D\xUsurf+\xWexp.
\eqe

The expansion work $\xWexp$ executed during the transformation can easily be calculated as minus the compression work received by the liquid at the constant pressure $\xpL$, whose volume decrease is the same as the vapour volume increase, thus obtaining
\eqb
\xWexp=\xpL(\xVc-\xVliq), \label{eq:5}
\eqe
where $\xVc$ is the final volume of the critically-sized vapour bubble, and $\xVliq$ is the initial volume containing the quantity of liquid which is to become the critically-sized vapour bubble such that $\xrhoL\xVliq = \xrhoV\xVc$.
This means that, if the internal energy variation $\D\xUvol$ from the metastable liquid state to the stable saturated vapour state is approximated using the difference between the internal energies of the stable saturated vapour and liquid states at pressure $\xpL$, then, according to the definition of the latent heat of vaporization based on the first law of thermodynamics, the sum $\D\xUvol + \xWexp$ equals the heat $\xQevap$ required for the phase change to occur at the constant pressure $\xpL$ \cite{bib:31}, i.e.,
\eqb
\xEc=\xQevap+\D\xUsurf. \label{eq:6}
\eqe

The same conclusion can be achieved by simply considering that the injection of $\xEc$ at constant pressure $\xpL$ results in an enthalpy variation $\D\xH$, which can be written as the sum of a volume term, $\D\xHvol$, and a surface term, $\D\xHsurf$. The $\D\xHvol$ term can be approximated by the heat required for the phase change at the liquid pressure $\xQevap$, whereas the $\D\xHsurf$ term, on account of the definition of enthalpy as $\xH=\xU+\xp\xV$, coincides with the internal energy variation $\D\xUsurf$, since the bubble surface has no volume. The heat of vaporization is given by
\eqb
\xQevap=\fb{4}{3}\pi\xRc^{3}\xrhoV\xlam \label{eq:7}
\eqe
in which, therefore, both $\xrhoV$ and $\xlam$ have to be evaluated at the stable equilibrium temperature at which the phase change takes place at the constant pressure $\xpL$, that is to say, the saturation temperature $\xTV$.

The internal energy variation of the bubble surface $\D\xUsurf$ can be calculated considering that the energy required to form the bubble surface is expressed in terms of the free energy, whose variation associated with the formation of a unit surface area equals the surface tension of the liquid, $\xsig=\fs{\d\xF}{\d\xA}$. Therefore, on account of the definition of free energy as $\xF=\xU-\xT\xS$, the internal energy variation consequent to the formation of the bubble surface at the constant temperature $\xTL$ can be written as
\eqb
\D\xUsurf=\D\xFsurf+\xTL\D\xSsurf, \label{eq:8}
\eqe
in which the variation of any state function clearly coincides with the value of the state function at the end of the bubble formation. According to the first law of thermodynamics, if \eqref{eq:8} is rewritten as $\D\xUsurf=\xTL\D\xSsurf - (-\D\xFsurf)$, then $\xTL\D\xSsurf$ represents the heat that must be supplied to the bubble surface to keep it at the constant temperature $\xTL$, whereas $-\D\xFsurf$ is the isothermal work done by the bubble surface during its formation, or better, $\D\xFsurf$ is the work that must be supplied to the bubble surface to allow its formation. 
The free energy variation $\D\xFsurf$ is given by the product of the area of the bubble surface multiplied by the surface tension
\eqb
\D\xFsurf=4\pi \xRc^{2}\xsig. \label{eq:9}
\eqe
The entropy variation $\D\xSsurf$, computed in terms of the entropy of the bubble surface at the end of its formation $\xSsurf$ is given by minus the temperature derivative of the surface free energy $\xFsurf$ calculated using \eqref{eq:9}.
In fact, on account of the first and second laws of thermodynamics, the differential $\d\xF$ is
\eqb
\d\xF=-\xp\d\xV-\xS\d\xT, \label{eq:10}
\eqe
suggesting
\eqb
\xF=\xF(\xV,\xT), \label{eq:11}
\eqe
\eqb
\d\xF=\(\fb{\de\xF}{\de\xV}\)_{\xT}\d\xV+\(\fb{\de\xF}{\de\xT}\)_{\xV}\d\xT, \label{eq:12}
\eqe
\eqb
-\(\fb{\de\xF}{\de\xV}\)_{\xT}=\xp, \label{eq:13}
\eqe
\eqb
-\(\fb{\de\xF}{\de\xT}\)_{\xV}=\xS. \label{eq:14}
\eqe
Notice that, since the bubble surface has no volume, the free energy of the bubble surface is a function of the temperature only, which results in $\xSsurf = -\fs{\d\xFsurf}{\d\xT}$, thus implying
\eqb
\D\xSsurf=-4\pi\xRc^{2}\fb{\d\xsig}{\d\xT}. \label{eq:15}
\eqe
The combination of \eqref{eq:8}, \eqref{eq:9} and \eqref{eq:15} gives
\eqb
\D\xUsurf=4\pi\xRc^{2}\(\xsig -\xTL\fb{\d\xsig}{\d\xT}\). \label{eq:16}
\eqe
Of course, the same relation \eqref{eq:16} can also be achieved using other ways, see, e.g., \cite{bib:32}. Substituting \eqref{eq:7} and \eqref{eq:16} in \eqref{eq:6}, we obtain
\eqb
\xEc=\fb 4 3\pi\xRc^{3}\xrhoV\xlam +4\pi\xRc^{2}\(\xsig -\xTL\fb{\d\xsig}{\d\xT}\). \label{eq:17}
\eqe

Accordingly, neither the expansion term, nor the kinetic energy term have to be included in the expression of the critical energy. In particular, as seen earlier, the expansion work done by the vapour bubble during its formation is already comprised in the vaporization term. As a matter of fact, the vaporization term consists of both the energy required to break the intermolecular bonds in the liquid, which results in an increased internal energy of the vapour phase, and the energy required to draw the vapour molecules apart, which corresponds to the positive expansion work transferred to the liquid. Thus, all the authors who add the expansion term in the critical energy equation assume that the heat required for the evaporation of the liquid is responsible only for the volume internal energy increase. On the other hand, those authors who subtract the expansion term simply proceed with the calculation of the volume internal energy variation instead of the heat required for the phase change. Finally, the kinetic energy imparted by the expanding vapour bubble to the surrounding liquid is nothing more than the same expansion work transferred from the vapour bubble to the surrounding liquid as perceived by the liquid, which implies that, as observed earlier for the expansion work term, also the kinetic energy term needs not to be considered.

At this stage, although not strictly required, it seems interesting to mention a consideration on the computation of the work executed during the bubble formation $\xWb$ in the hypothesis of reversibility of the transformation
\eqb
\xWb=\int_{b}\xpb\d\xV, \label{eq:18}
\eqe
where the equilibrium pressure inside the vapour bubble $\xpb$ can be directly derived from \eqref{eq:2} by simply replacing $\xRc$ with $\xr$, and the infinitesimal volume variation $\d\xV$ can be expressed as $4\pi \xr^{2}\di\xr$, thus following
\eqb
\xWb=\int_{b}\(\xpL+\fb{2\xsig }{\xr}\)4\pi \xr^{2}\di\xr. \label{eq:19}
\eqe
Hence
\eqb
\xWb=\int_{\text{vol}}\xpL 4\pi\xr^{2}\di\xr + \int_{\text{surf}} 8 \xsig \pi \xr\di\xr, \label{eq:20}
\eqe
which points out that the work done during the vapour bubble formation is actually composed of a volume term, $\xWvol$, and a surface term, $\xWsurf$. Taking into account that the liquid pressure $\xpL$ is constant, and assuming that the surface tension $\xsig$ is substantially independent of the vapour bubble curvature \cite{bib:17}, after some algebra we obtain
\eqb
\xWb=\xWvol+\xWsurf=\fb 4 3 \pi \xRc^{3}\xpL\(1-\fb{\xrhoV}{\xrhoL}\)+4\pi\xRc^{2}\xsig, \label{eq:21}
\eqe
in which the volume term can be identified as the expansion work transferred from the vapour bubble to the surrounding liquid $\xWexp$ given by \eqref{eq:5}, whereas, based on \eqref{eq:9}, the surface term is the work done to form the vapour bubble surface, which therefore is not transferred out of the bubble.

\section{Radius of the critically-sized nucleated vapour bubble}
\begin{figure}[t!]
\centering
\includegraphics[width=3.5772in,height=2.6071in]{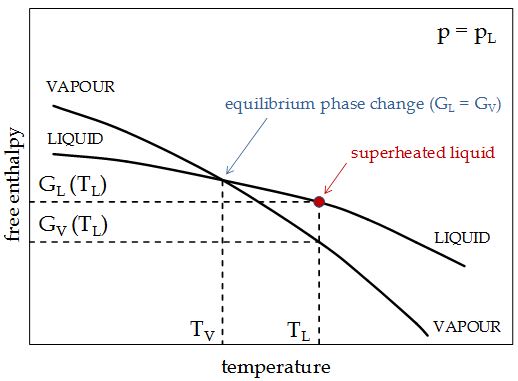} 
\caption{Distributions of $\xGL$ and $\xGV$ vs. $\xT$ at $\xp = \xpL$}
\label{fig:2}
\end{figure}

The radius $\xRc$ of the critically-sized vapour bubble given by \eqref{eq:2} is normally calculated in the hypothesis of stable equilibrium conditions, despite this is not the real situation. Indeed, pressure $\xpb$ is usually approximated using the saturation pressure at the liquid temperature \cite{bib:8,bib:9,bib:10,bib:11,bib:13,bib:14,bib:15,bib:16,bib:18,
bib:20,bib:22,bib:23,bib:24,bib:25,bib:26,bib:27,bib:28,bib:29,bib:30}, which gives
\eqb
\xRc\approx \fb{2\xsig }{\xpV-\xpL}. \label{eq:22}
\eqe
Differently, some authors \cite{bib:3,bib:12,bib:17,bib:19,bib:21} approximate $\xpb$ using the pressure value obtained by imposing the customary stable equilibrium condition of equality of the chemical potentials, or, that is the same, the specific free enthalpies of the metastable liquid and the stable vapour at the liquid temperature, and assuming that the mass densities of the liquid and vapour phases are substantially the same as their corresponding saturation values at the liquid temperature, which results in
\eqb
\xRc\approx \fb{2\xsig }{\(\xpV-\xpL\)\(1-\fb{\xrhoV}{\xrhoL}\)}. \label{eq:23}
\eqe
Actually, although both mentioned approximations can be considered as reasonably true at low degrees of metastability, their application at the high superheats asked for WIMP-recoil detection can lose accuracy. In this regard, an alternative approach can be followed by recalling that, when a thermodynamic system kept at constant temperature and pressure can be in more than one equilibrium state, then the stable equilibrium state is the state of lowest free enthalpy, also named Gibbs free energy, and defined as $\xG=\xH-\xT\xS$, which therefore plays the same role played by the potential energy in defining the stable equilibrium state of a mechanical system \cite{bib:31}. It follows that in the present case the free enthalpy of the superheated liquid $\xGL$ is necessarily higher than that of the stable vapour $\xGV$ of an amount $\D\xG$ that can also be seen as the free enthalpy variation associated with the formation of a vapour bubble. The situation is schematically displayed in figure \ref{fig:2}, where typical distributions of $\xGL$ and $\xGV$ at the constant pressure $\xpL$ are plotted versus $\xT$. It is apparent that, since $\D\xG$ increases as the metastability degree is increased, the assumption of equality of $\xGL$ and $\xGV$ becomes inaccurate at high degrees of superheat.
Thus, a more realistic approach is required which should be able to reflect that the critical size represents a condition of absolute instability for the vapour bubble. In fact, should the critically-sized vapour bubble lose just a tiny amount of matter, say one molecule, which gets back to be part of the surrounding liquid, then the bubble will literally implode, vanishing, due to the loss of the mechanical equilibrium. Conversely, should the critically-sized vapour bubble gain just a tiny amount of matter, taken away from the surrounding liquid, then the bubble will spontaneously grow, becoming detectable.

In view of the mentioned relation between the value of the free enthalpy and the stability of a system that can be in more than one equilibrium state at constant $\xT$ and $\xp$, such an extreme instability condition must correspond to a maximum of the difference between the free enthalpies of the superheated liquid and the stable vapour, or, that is the same, a maximum of the free enthalpy variation associated with the phase change, which is an approach also used in the study of crystal nucleation, see, e.g., \cite{bib:33,bib:34,bib:35,bib:36}. The critical radius $\xRc$ can then be regarded as the size of the vapour bubble corresponding to the maximum of the function which describes the free enthalpy variation $\D\xG(\xr)$ associated with the formation of a vapour bubble of radius $\xr$ that nucleates in a metastable liquid kept at constant temperature $\xTL$ and pressure $\xpL$
\eqb
\D\xG(\xr)=\D\xH(\xr)-\xTL\D\xS(\xr), \label{eq:24}
\eqe
which can also be written as the sum of a volume term $\D\xGvol(\xr)$ and a surface term $\D\xGsurf(\xr)$ giving
\eqb
\D\xG(\xr)=\D\xGvol(\xr)+\D\xGsurf(\xr). \label{eq:25}
\eqe
The volume term can be expressed as
\eqb
\D\xGvol(\xr)=\D\xHvol(\xr)-\xTL\D\xSvol(\xr), \label{eq:26}
\eqe
where, as previously done for the internal energy variation, the enthalpy and entropy variations from the metastable liquid state to the stable saturated vapour state can be approximated using the respective stable equilibrium variations at temperature $\xTV$, provided that the superheat degree is sufficiently small compared with the difference between the critical and triple points. Actually, this is an easy way to estimate $\D\xHvol(\xr)$ and $\D\xSvol(\xr)$, whose values would be otherwise difficult to determine, and to account for the metastability degree in the derivation of $\D\xG(\xr)$. Therefore, $\D\xHvol(\xr)$ and $\D\xSvol(\xr)$ are calculated as the heat required for the phase change at temperature $\xTV$, and the heat required for the phase change at temperature $\xTV$ divided by the same temperature $\xTV$, respectively
\eqb
\D\xHvol(\xr)=\fb 4 3\pi\xr^{3}\xrhoV\xlam, \label{eq:27}
\eqe
\eqb
\D\xSvol(\xr)=\fb{\fb 4 3\pi\xr^{3}\xrhoV\xlam}{\xTV}, \label{eq:28}
\eqe
in which both $\xrhoV$ and $\xlam$ must be evaluated at temperature $\xTV$. Hence
\eqb
\D\xGvol(\xr)=\fb{4}{3}\pi\xr^{3}\xrhoV\xlam\(1-\fb{\xTL}{\xTV}\). \label{eq:29}
\eqe

As far as the surface term is concerned, based on the cited definitions of enthalpy and free energy, the free enthalpy can be expressed as $\xG=\xF+\xp\xV$, thus following that, since the bubble surface has no volume, the free enthalpy change $\D\xGsurf(\xr)$ coincides with the free energy change $\D\xFsurf(\xr)$, which can be directly derived from \eqref{eq:9} by simply replacing $\xRc$ with $\xr$, i.e.,
\eqb
\D\xGsurf(\xr)=4\pi\xr^{2}\xsig. \label{eq:30}
\eqe
The combination of \eqref{eq:25}, \eqref{eq:29} and \eqref{eq:30} gives
\eqb
\D\xG(\xr)=-\fb 4 3\pi\xr^{3}\xrhoV\xlam \fb{\xTL-\xTV}{\xTV}+4\pi\xr^{2}\xsig. \label{eq:31}
\eqe
Indeed, \eqref{eq:31} can also be obtained by determining $\D\xH(\xr)$ and $\D\xS(\xr)$, and substituting their expressions in \eqref{eq:24}.

The enthalpy variation $\D\xH(\xr)$, equal to the heat injection required to nucleate the bubble, can be directly derived from \eqref{eq:17}, by simply replacing $\xRc$ with r, i.e., 
\eqb
\D\xH(\xr)=\fb 4 3\pi\xr^{3}\xrhoV\xlam +4\pi\xr^{2}\(\xsig -\xTL\fb{\d\xsig}{\d\xT}\), \label{eq:32}
\eqe
where, like before, $\xrhoV$ and $\xlam$ must be evaluated at temperature $\xTV$; conversely, the values of $\xsig$ and d$\fs{\xsig}{\d\xT}$ are referred to temperature $\xTL$.

On the other hand, the entropy variation $\D\xS(\xr)$ can be written as the sum of a volume term, $\D\xSvol(\xr)$, and a surface term, $\D\xSsurf(\xr)$. The volume term $\D\xSvol(\xr)$ is given by \eqref{eq:28}, whereas the surface term $\D\xSsurf(\xr)$ can be directly derived from \eqref{eq:15}, by simply replacing $\xRc$ with r, thus obtaining
\eqb
\D\xS(\xr)=\fb{\fb 4 3\pi\xr^{3}\xrhoV\xlam }{\xTV}-4\pi\xr^{2}\fb{\d\xsig}{\d\xT}, \label{eq:33}
\eqe
in which both $\xrhoV$ and $\xlam$ must be evaluated at temperature $\xTV$, while $\fs{\d\xsig}{\d\xT}$ has to be calculated at temperature $\xTL$.

\begin{figure}[t!]
\centering
\includegraphics[width=3.6965in,height=2.6457in]{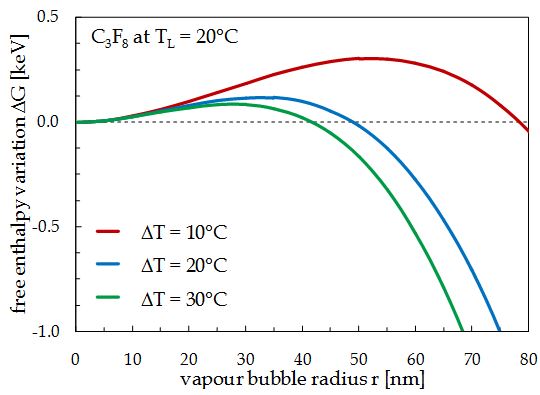} 
\caption{Distributions of $\D\xG(\xr)$ vs. $\xr$ for C$_{3}$F$_{8}$ at $\xTL$ = 20 $^{\circ}$C, using $\D\xT$ as a parameter}
\label{fig:3}
\end{figure}

The critical radius of the vapour bubble, $\xRc$, is then determined by computing the root of the derivative $\D\xGprime(\xr)$, again assuming that $\xsig$ is independent of the vapour bubble curvature \cite{bib:17}, which results in
\eqb
\xRc=\fb{2\xsig }{\xrhoV\xlam \fb{\xTL-\xTV}{\xTV}}. \label{eq:34}
\eqe
Hence, the formation of a vapour bubble occurs via a pathway involving the surmounting of the barrier of potential $\D\xG(\xRc)$, whose value is given by \eqref{eq:31} with $\xr = \xRc$. A number of distributions of $\D\xG(\xr)$ relative to C$_{3}$F$_{8}$, i.e., the target liquid used for WIMP-recoil detection in the experiments carried out by PICO \cite{bib:6} and MOSCAB \cite{bib:7}, are plotted in figure \ref{fig:3} against the radius $\xr$ for $\xTL$ = 20 $^{\circ}$C using the superheat degree $\D\xT$ as a parameter. The values of the physical properties are taken from the NIST Chemistry WebBook \cite{bib:37}. 

Of course, should the degree of metastability of the superheated liquid be sufficiently low, then \eqref{eq:1} can be rewritten by approximating the temperature derivative $\fs{\d\xp}{\d\xT}$ with the corresponding increment ratio, i.e., the ratio between the underpressure $\D\xp$ and the superheat $\D\xT$, thus obtaining
\eqb
\fb{\xpV-\xpL}{\xTL-\xTV}\approx \fb{\xrhoV\xlam }{\xTV(1-\xrhoV/\xrhoL)} \label{eq:35}
\eqe
and then
\eqb
\xrhoV\xlam \fb{\xTL-\xTV}{\xTV}\approx (\xpV-\xpL)\(1-\fb{\xrhoV}{\xrhoL}\), \label{eq:36}
\eqe
in which, due to the low $\D\xT$, the density ratio at temperatures $\xTV$ and $\xTL$ is practically the same. The replacement of \eqref{eq:36} in \eqref{eq:34} leads to \eqref{eq:23}. Moreover, if we take into account that $\fs{\xrhoV}{\xrhoL} \ll 1$, \eqref{eq:36} can be further reduced to 
\eqb
\xrhoV\xlam \fb{\xTL-\xTV}{\xTV}\approx \xpV-\xpL, \label{eq:37}
\eqe
which, replaced in \eqref{eq:34}, leads to \eqref{eq:22}.

\begin{figure}[t!]
\centering
\includegraphics[width=3.6945in,height=2.6291in]{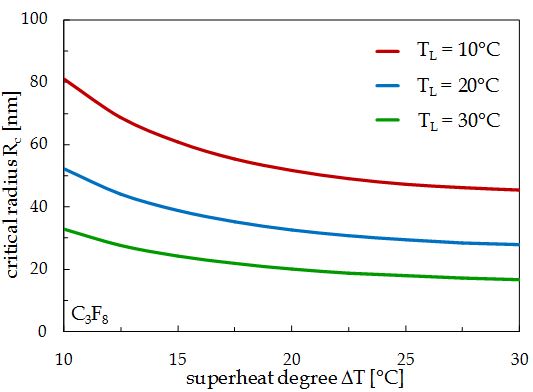} 
\caption{Distributions of $\xRc$ vs. $\D\xT$ for C$_{3}$F$_{8}$ using $\xTL$ as a parameter}
\label{fig:4}
\end{figure}

\begin{figure}[t!]
\centering
\includegraphics[width=3.6992in,height=2.6346in]{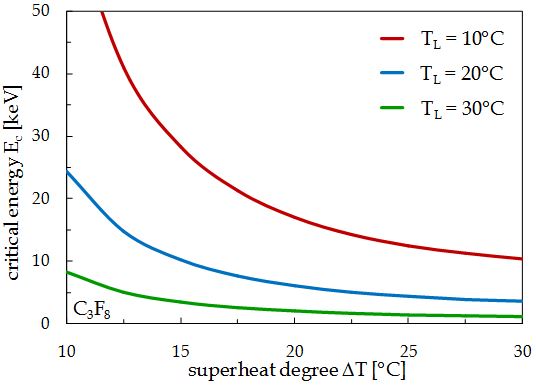} 
\caption{Distributions of $\xEc$ vs. $\D\xT$ for C$_{3}$F$_{8}$ using $\xTL$ as a parameter}
\label{fig:5}
\end{figure}

Two sets of distributions of the critical radius expressed by \eqref{eq:34} and the related critical energy expressed by \eqref{eq:17}, plotted against the superheat degree using the liquid temperature as a parameter, are reported in figures \ref{fig:4} and \ref{fig:5} for C$_{3}$F$_{8}$. 

\section{Discussion}
First of all, it is worth observing that the procedure followed to obtain \eqref{eq:34} by determining $\D\xH(\xr)$ and $\D\xS(\xr)$, and then substituting their expressions in \eqref{eq:24}, intrinsically demonstrates the validity of \eqref{eq:17}. In fact, should the heat injection required to nucleate a vapour bubble have been derived from a relationship different from \eqref{eq:17}, then a relationship different from \eqref{eq:34} would have been achieved for the critical radius $\xRc$, and neither \eqref{eq:23} nor \eqref{eq:22} could have been obtained for low degrees of metastability.

Another point worth being quoted is that, although usually either no mention is done on the temperature at which the physical properties have to be evaluated or explicit reference is made to the temperature of operation, just the surface tension and its temperature derivative have to be calculated at the liquid temperature $\xTL$, while, based on the approach discussed above, both $\xrhoV$ and $\xlam$ should be evaluated at the vapour temperature $\xTV$.

\begin{figure}[t!]
\centering
\includegraphics[width=3.6972in,height=2.6661in]{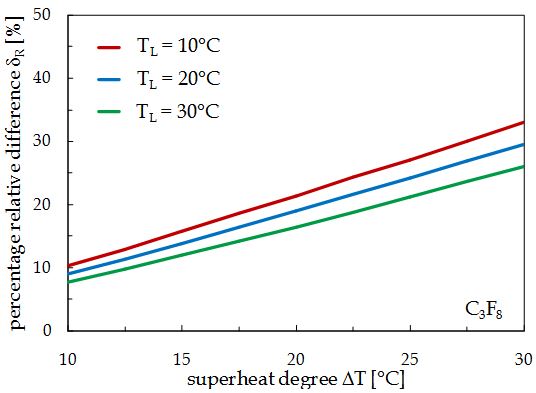} 
\caption{Distributions of $\xdelR = \fs{(\xRc-\xRcstar)}{\xRc}$ vs. $\D\xT$ for C$_{3}$F$_{8}$ using $\xTL$ as a parameter}
\label{fig:6}
\end{figure}

Furthermore, it must be pointed out that the calculation of the critical radius $\xRc$ by the way of \eqref{eq:22} or \eqref{eq:23} leads to values lower than that expressed by \eqref{eq:34}, which is a direct consequence of the fact that, since the vapour pressure curve is concave upwards, the temperature derivative of the saturation pressure at temperature $\xTV$ is lower than the corresponding increment ratio $\fs{(\xpV-\xpL)}{(\xTL-\xTV)}$. Of course, the discrepancy increases as the degree of metastability is increased, as shown in figure \ref{fig:6}, in which a number of distributions of the relative difference $\xdelR=\fs{(\xRc-\xRcstar)}{\xRc}$ between the results obtained applying \eqref{eq:23} instead of \eqref{eq:34} are plotted against the superheat degree $\D\xT$ for C$_{3}$F$_{8}$ using the liquid temperature $\xTL$ as a parameter, where $\xRc$ and $\xRcstar$ are the values of the critical radius given by \eqref{eq:34} and \eqref{eq:23}, respectively. Even higher discrepancies are obtained if \eqref{eq:22} is applied rather than \eqref{eq:23}.

\begin{figure}[t!]
\centering
\includegraphics[width=3.6957in,height=2.6717in]{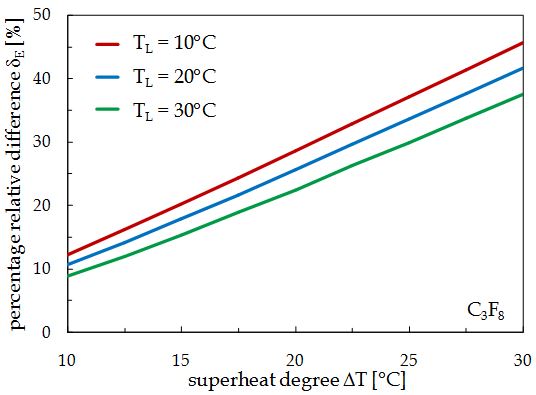} 
\caption{Distributions of $\xdelE = \fs{(\xEc - \xEcstar)}{\xEc}$ vs. $\D\xT$ for C$_{3}$F$_{8}$ using $\xTL$ as a parameter}
\label{fig:7}
\end{figure}

Accordingly, the critical energy obtained through \eqref{eq:17} in which $\xRc$ is calculated by \eqref{eq:34} is higher than the critical energy derived applying, for example, the equation proposed by Bugg \cite{bib:10} (third line of table \ref{tab:1}) using \eqref{eq:23} to calculate $\xRc$. In addition, the absence of the subtractive expansion term, contribute to the further increase of the value of the critical energy.
A set of distributions of the relative difference $\xdelE = \fs{(\xEc - \xEcstar)}{\xEc}$ between the results obtained applying the Bugg's equation in combination with \eqref{eq:23}, instead of \eqref{eq:17} in combination with \eqref{eq:34}, are plotted in figure \ref{fig:7} against the superheat degree $\D\xT$ for C$_{3}$F$_{8}$ using the liquid temperature $\xTL$ as a parameter, where $\xEc$ and $\xEcstar$ are the values of the critical energy given by \eqref{eq:17} and by the Bugg's equation, respectively. It is apparent that when the degree of metastability of the superheated liquid is high enough, the relative difference between the two values becomes significantly high whatever the liquid temperature is, which may notably affect the estimation of the bubble chamber sensitivity. 

As a matter of fact, the computation of the heat required for the phase change executed at temperature $\xTV$, instead of $\xTL$, is the main responsible for the higher critical energies obtained using \eqref{eq:17} and \eqref{eq:34}, whose application may thus be regarded as a prudential approach to the problem, resulting in what we could call an upper theoretical limit of the thermodynamic energy threshold.

Finally, it seems interesting to compare the theoretical prediction of the combination of \eqref{eq:17} and \eqref{eq:34} with the experimental result recently obtained for liquid Xenon, which is definitely more indicated than the other usual target liquids to validate a novel critical energy theoretical equation. Actually, at any recoil energy, an ion of Xenon travelling in pure liquid Xenon has a so much higher stopping force than, for example, $^{12}$C or $^{19}$F in liquid C$_{3}$F$_{8}$, or other carbon-fluorine compounds, to more closely match the thermal spike theory. On the other hand, the use of a single-atom target gives rise to less uncertainties in defining the threshold. In fact, when a multi-component substance is used, a single calibration point (corresponding to a measured bubble rate produced at a given pressure and temperature condition by a single spectrum of nuclear recoil energies) can be fit by several sets of efficiency curves, which means that, should the critical energy be underestimated, and the related response of the heaviest ion be overestimated, the same bubble rate can be accomplished through a fit assuming a much smaller contribution from the lighter components. 

The cited threshold measurement was performed by Baxter et al. \cite{bib:38} using a 30-g Xenon bubble chamber operated at 30 psia and $-$60 $^{\circ}$C, whose corresponding critical energy calculated by the Bugg's equation in combination with \eqref{eq:23} would be 8.3 keV. Indeed, the observed single and multiple bubble rates consequent to a 3.1 h exposure to a $^{252}$Cf \ neutron source were consistent with the absolute rates predicted by a Monte Carlo simulation of the equipment executed using the MCNPX-POLIMI package assuming that the minimum nuclear recoil energy required to nucleate a vapour bubble was 19 $\pm$ 6 keV (i.e., more than the double of 8.3 keV), where, according to the authors, the range was dominated by the 30\% uncertainty in their source strength. Conversely, the application of the relationships proposed for the calculation of $\xEc$ and $\xRc$, i.e., \eqref{eq:17} and \eqref{eq:34}, results in a theoretical value of the critical energy equal to 20.2 keV.

\section{Conclusions}
The relationships currently available for the calculation of the critical energy required for homogeneous nucleation in a superheated liquid $\xEc$ and the corresponding critical radius of the nucleated vapour bubble $\xRc$ show a number of inconsistencies. This is the reason why their application may result in a more or less noticeable overestimation of the sensitivity of the bubble chambers employed in dark matter searches, and also affect the equipment calibration, which has motivated the present study. Actually, based on the procedure followed to obtain them, the pair of equations proposed here for the calculation of $\xEc$ and $\xRc$ turn out to be more consistent with the physical facts, the first being based on the application of the first law of thermodynamics, the second being derived under the assumption that the extreme instability condition represented by the critically-sized vapour bubble must correspond to a maximum of the difference between the free enthalpies of the metastable liquid and the stable vapour phases. A good agreement has also been found between our theoretical prediction and an experimental result recently reported for Xenon at 30 psia and $-$60 $^{\circ}$C. Further investigations on this topic are scheduled to be conducted in the next future.

\acknowledgments
The authors are grateful to Donald Cundy for the valuable discussions and suggestions and for his help in reviewing the manuscript.

\end{document}